\documentclass[twocolumn,final,aps,prb,showpacs,amsmath,amssymb,amsfonts,floatfix]{revtex4-1}

\usepackage{url}

\usepackage{amsmath,amssymb,amsfonts}
\usepackage{color}
\usepackage[colorlinks,breaklinks,bookmarks=true,citecolor=blue,linkcolor=red,urlcolor=blue]{hyperref}
\definecolor{darkred}{rgb}{0.7,0.0,0}

\usepackage{graphicx}
\usepackage{multirow}
\usepackage{epsfig}

\begin{document}
\title{Nickelate superconductors -- a renaissance of the one-band Hubbard model}


\author{Motoharu Kitatani$^{1,2,{*}}$, Liang Si$^{1,3,{*}}$,  Oleg Janson$^{4}$, Ryotaro Arita$^{2,5}$,  Zhicheng Zhong$^{3}$,  Karsten Held$^{1,\dagger}$}
\affiliation{$^1$ Institute for Solid State Physics, Vienna University of Technology, 1040 Vienna, Austria \\  $^2$  RIKEN Center for Emergent Matter Sciences (CEMS), Wako, Saitama, 351-0198, Japan
 \\  $^3$ Key Laboratory of Magnetic Materials and Devices \& Zhejiang Province Key Laboratory of Magnetic Materials and Application Technology, Ningbo Institute of Materials Technology and Engineering (NIMTE), Chinese Academy of Sciences, Ningbo 315201, China \\  $^4$ Leibniz Institute for Solid State and Materials Research IFW Dresden, 01171 Dresden, Germany \\  $^5$  Department of Applied Physics, The University of Tokyo, Hongo, Tokyo, 113-8656, Japan}
\date{\today}

  \newenvironment{methods}{\section*{Methods}\appendix}{}
  \newenvironment{addendum}{\section*{Addendum}\begin{itemize}}{\end{itemize}}

\maketitle

{\bf
Following the  discovery of superconductivity in the cuprates\cite{Bednorz1986} and the seminal work by Anderson\cite{Anderson1987}, the theoretical efforts to understand  high-temperature superconductivity have been focusing to a large extent on a simple model: the one-band Hubbard model\cite{Editorial2013,Gull2015,Jiang2019b}. However,  superconducting  cuprates need to be doped, and the doped holes go into the oxygen orbitals\cite{Zaanen1985,Fujimori1987,Pellegrin1993}. This requires a more elaborate multi-band model such as the three-orbital Emery model\cite{Emery1987}.\footnote{If at all, the Hubbard model may mimic the physics of the  Zhang-Rice singlet [F.\,C.~Zhang and T.\,M.\,~Rice, Phys.\ Rev.\ B {\bf 37}, {3759} (1988)] between the moment on the copper sites and the oxygen holes.} The recently discovered nickelate superconductors\cite{li2019superconductivity} appear, at first glance, to be even more complicated multi-orbital systems. Here, we analyse this multi-orbital system and find that it is instead the nickelates which can be described by a one-band Hubbard model, albeit with an additional electron reservoir and only around the superconducting  regime. 
 Our calculations of the critical temperature ${\mathbf T_{\rm C}}$ are in good agreement with experiment, and show that optimal doping is slightly below the 20\% Sr-doping of Ref.~\onlinecite{li2019superconductivity}. Even more promising than  3$d$ nickelates are 4${\mathbf \it d}$ palladates.
}

The discovery of superconductivity in Sr$_{0.2}$Nd$_{0.8}$NiO$_2$ by Li {\em et al.}\cite{li2019superconductivity} marked the beginning of a new, a nickel age of superconductivity, ensuing  a plethora of experimental and theoretical work; see, among others, Refs.~\onlinecite{Li2019,Zhou2019,Lee2020,Botana2019,Hirofumi2019,Motoaki2019,hu2019twoband,Wu2019,Nomura2019,Zhang2019,Gao2019,Jiang2019,Liu2019,Ryee2019,Lechermann2019,Zhang2019b,Si2020,Werner2020,Geisler2020,Bernardini2020,Choi2020,Karp2020}. Similar as for the cuprates, the basic structural elements are  NiO$_2$ planes on a square lattice, and Ni   has the same  formal 3$d^9$  electronic configuration.

\begin{figure}[tb]
\includegraphics[width=8.8cm]{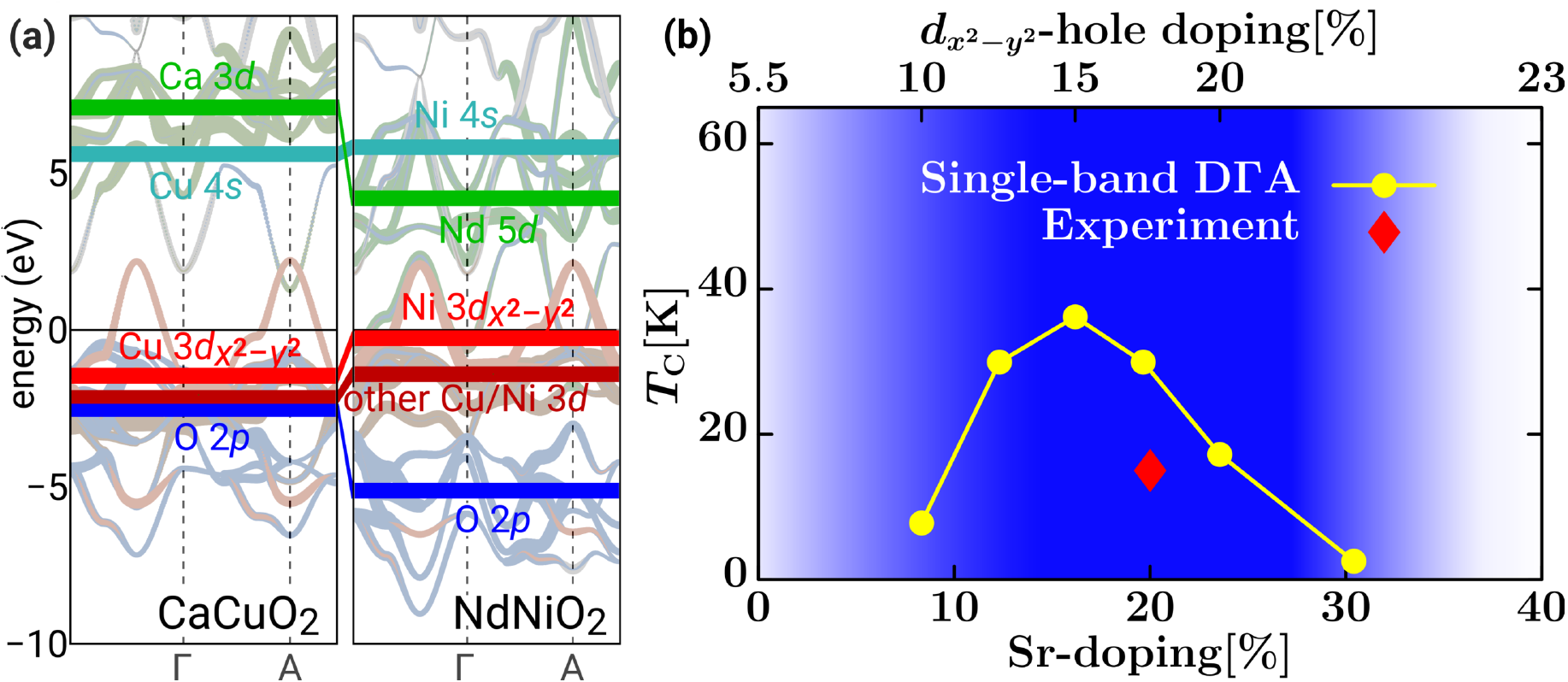}
%
\caption{(a) Energy levels  comparing  CaCuO$_2$  and NdNiO$_2$, with the DFT bandstructure as a background.
(b) Phase diagram $T_{\rm C}$ vs.~Sr-doping as calculated in D$\Gamma$A together with the hitherto only experimental point:  15\,K for 20\% Sr-doping\cite{li2019superconductivity}. In the blue-shaded region, a one-band Hubbard model description is possible with its  doping given on the upper $x$-axis. \label{Fig:overview}}
\end{figure}

But at second glance, there are noteworthy differences, see Fig.~\ref{Fig:overview}\,(a). For the parent compound NdNiO$_2$,  density functional theory (DFT) calculations show, besides the Ni $d_{x^2-y^2}$ orbital, additional bands around the Fermi level $E_F$ \cite{Lee2004,Nomura2019,Zhang2019,Botana2019} that are of predominant Nd-5$d$ character and overlap with the former.  Note that  the Nd-5$d$ bands in  Fig.~\ref{Fig:overview}\,(a) extend below $E_F$  despite their  centre of gravity being considerably above $E_F$.

Such bands or electron pockets have the intriguing effect that even for the parent compound  NdNiO$_2$, the Ni-$d_{x^2-y^2}$ orbital is hole doped, with the missing electrons in the Nd-5$d$ pockets. In other words, the parent compound  NdNiO$_2$ already behaves as the doped cuprates. First calculations\cite{Hirofumi2019,Wu2019,Zhang2019b} for  superconductivity in the nickelates  that are valid at weak interaction strength hence started from a Fermi surface with both, the  Nd-5$d$ pockets and the  Ni-$d_{x^2-y^2}$ Fermi surface. Such a multi-orbital nature of superconductivity has also been advocated in Refs.~\onlinecite{hu2019twoband,Werner2020}.

There is further the Nd-4$f$ orbital which in density functional theory (DFT) spuriously shows up just above  the Fermi energy. But these  4$f$ orbitals will be localised which can be mimicked by DFT+$U$ or by putting them into the core, as has been done in Fig.~\ref{Fig:overview} (a). Choi {\em et al.}\cite{Choi2020} suggest a  ferromagnetic, i.e., anti-Kondo coupling of these  Nd-4$f$ with the aforementioned Nd-$5d$ states.

A further striking difference is that the oxygen band is much further away from the Fermi level than for the cuprates, see Fig.~\ref{Fig:overview} (a). Vice versa, the other Ni-3$d$ orbitals are closer to the Fermi energy and  slightly doped because of their hybridisation with the   Nd-5$d$ orbitals.  Hence, in dynamical mean-field theory (DMFT)\cite{RevModPhys.68.13} calculations albeit employing a rather large interaction so that NdNiO$_2$ is a Mott insulator,
Lechermann \cite{Lechermann2019} finds that the holes in doped Sr$_{0.2}$Nd$_{0.8}$NiO$_2$  go to a larger extent to the  Ni-$d_{z^2}$ orbital, but not the O-$p$ orbitals. In the extreme situation of one extra hole per Ni site, Ni has a  3$d^8$ configuration with only one hole in the   $d_{x^2-y^2}$-orbital and the other one in the  $d_{z^2}$ orbital forming a local spin-1, as realised e.g.~in 
 LaNiO$_2$H.\cite{Si2020}  Other DMFT calculations for 
the parent compounds NdNiO$_2$\cite{Karp2020} and  LaNiO$_2$\cite{Si2020}, instead report a predominately  Ni 3$d_{x^2-y^2}$ band plus  Nd-5$d$ pockets. This leads to the questions: Which orbitals are depopulated  if we dope NdNiO$_2$ with Sr? Is a multi-orbital description necessary for the actual superconducting compound  Sr$_{x}$Nd$_{1-x}$NiO$_2$? In which doping regime is Sr$_{x}$Nd$_{1-x}$NiO$_2$ superconducting at all? What is the upper limit for the superconducting transition temperature $T_{\rm C}$?

In this paper, we show  that, if we properly include electronic correlation by DMFT, up to a Sr-doping $x$ of about 30\%, the holes 
only depopulate the  Ni-3$d_{x^2-y^2}$ and Nd-5$d$ bands. Only for larger Sr-dopings, holes are doped into the other Ni-3$d$ orbitals, necessitating a multi-orbital description. The hybridisation between the Ni-3$d_{x^2-y^2}$ and  Nd-5$d$  orbitals as calculated from the DFT-derived Wannier Hamiltonian is vanishing. 
We hence conclude that up to a Sr-doping of around  30\% marked as dark blue in Fig.~\ref{Fig:overview}\,(b), a single Ni-3$d_{x^2-y^2}$   band description as in the one-band Hubbard model is possible.  However, because of the Nd-5$d$ pocket(s), which acts like a electron reservoir and otherwise hardly interacts, only part of  the  Sr-doping [lower $x$-axis  of Fig.~\ref{Fig:overview}\,(b)]  goes into the Ni-3$d_{x^2-y^2}$-band [upper $x$-axis], cf.~Supplementary Information  Section~S.2 for the functional dependence.  We also take small Sr-doping out of  the blue-shaded region since for such small dopings there is, besides the Nd-5$d$ $A$-pocket, the $\Gamma$ pocket which interacts with the Nd-4$f$ moments ferromagentically\cite{Choi2020} and might result in additional correlation effects. At larger doping and when including the Nd-interaction in DMFT the $\Gamma$-pocket is shifted above $E_F$, see Fig.~\ref{Fig:DMFT} below. 

 Fig.~\ref{Fig:overview}\,(b) further shows the superconducting critical temperature $T_{\rm C}$ of the thus derived and doped Hubbard model, calculated by a method that is appropriate in the strong coupling regime: the dynamical vertex approximation (D$\Gamma$A\cite{RMPVertex}).  The agreement with experiment is reasonable given that the experimental $T_{\rm C}$ can be expected to be lower because of  e.g.~impurity scattering and the theoretical one is somewhat overestimated\footnote{\label{FN}The back coupling of the particle-particle channel to the particle-hole and transversal particle-hole channel is not included. While a (tiny)  hopping $t_z$ in the $z$-direction is necessary to overcome the Mermin-Wagner theorem, somewhat larger   $t_z$'s may suppress $T_{\rm C}$, see 
R. Arita, K. Kuroki, and H. Aoki, J. Phys. Soc. Japn. {\bf 69}, 1181 (2000); 
M. Kitatani, N. Tsuji, and H. Aoki, Phys. Rev. B {\bf 95},
075109 (2017).}.

Let us now discuss these results in more detail. We start with a DFT calculation [cf.~Supplementary Information~Section~S.1] which puts the Nd-4$f$ orbitals just above $E_F$. But since their hybridisation with the  Ni-3$d_{x^2-y^2}$ orbital is weak\cite{jiang2019electronic}, $|V_{x^2-y^2,4f}|=25\,$meV, see  Supplementary Information~Section S.2, they will localise and not make a Kondo effect. This localisation can be described e.g.~by the spin-splitting in DFT+$U$, or by including the  Nd-4$f$ states in the core. It leaves us with a well defined window with just  five Nd-5$d$ and five  Ni-3$d$ around the Fermi energy. For these remaining ten orbitals we do a Wannier function projection  (see Supplementary Information Section~S.2) and subsequent DMFT calculation with constrained random phase approximation (cRPA) calculated inter-orbital interaction $U'= 3.10\,$eV ($2.00\,$eV) and Hund's exchange  $J=0.65\,$eV  (0.25\,eV) for Ni (Nd)\cite{Si2020}.

Fig.~\ref{Fig:DMFT} presents the calculated DMFT spectral function for these ten bands. Let us first concentrate on Sr$_{0.2}$Nd$_{0.8}$NiO$_2$ for which also the $\mathbf k$-resolved spectral function on the right hand side is shown. Clearly in DMFT there is a single, compared to the DFT strongly renormalized Ni-3$d$ band of $d_{x^2-y^2}$ character  crossing $E_F=0$, see the zoom in Fig.~\ref{Fig:DMFT} (e). Besides, there is also a pocket around the $A$-point of predominately Nd-5$d_{xy}$ character, but the $\Gamma$ pocket is shifted above $E_F$, cf.~Supplementary Information Section~S.4 for other dopings. 

\begin{figure*}[t]
\includegraphics[width=17cm]{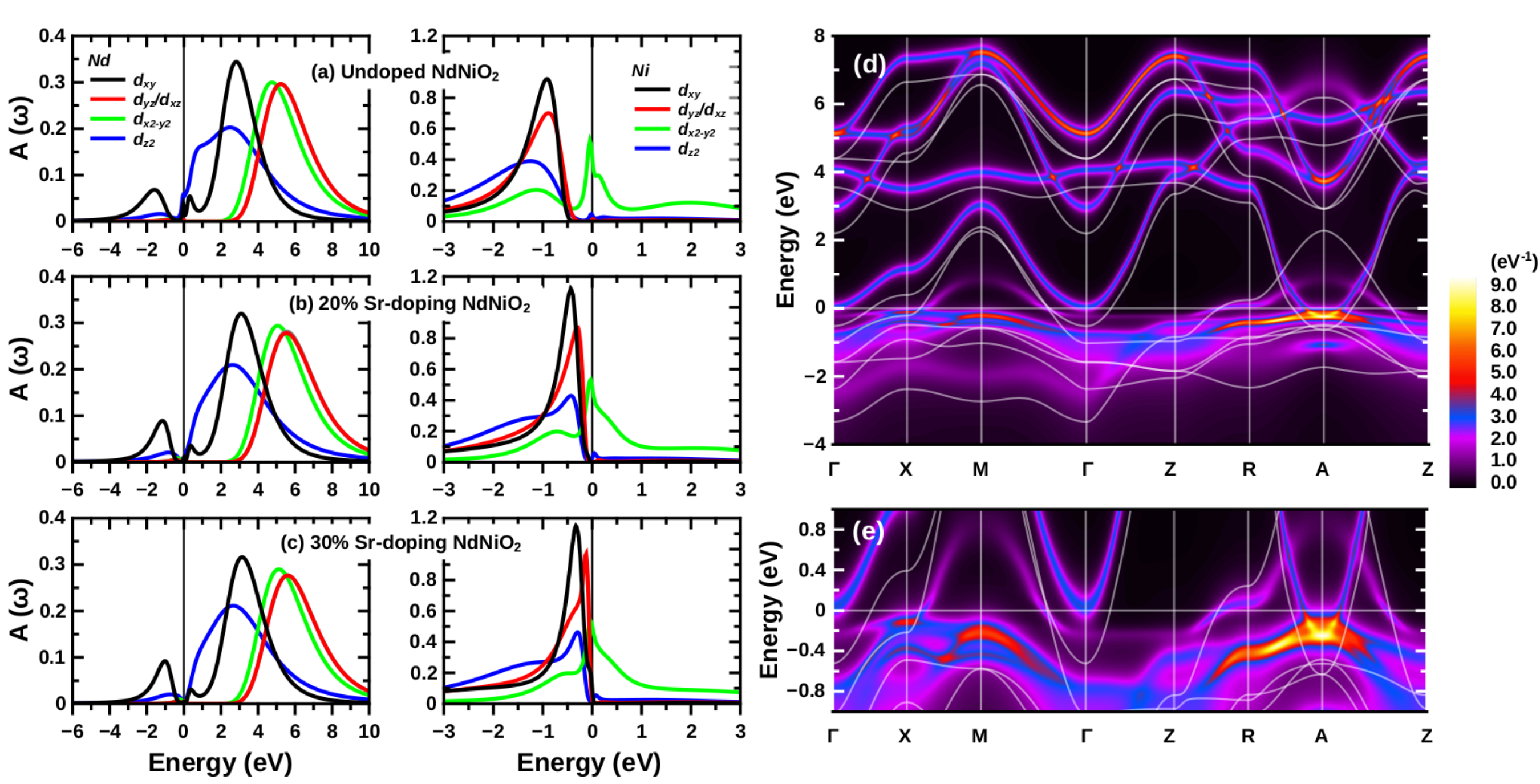}
\caption{(a-c) DFT+DMFT $\mathbf k$-integrated spectral function  $A(\omega)$
of Sr$_x$Nd$_{1-x}$NiO$_2$ at 0\%, 20\% and 30\% Sr-doping, orbitally resolved for the Nd-5$d$ and Ni-3$d$ orbitals. (d)  DFT+DMFT $\mathbf k$-resolved spectral function  $A(\omega)$ for  Sr$_{0.2}$Nd$_{0.8}$NiO$_2$. (e) Zoom-in of (d). The DFT bands are shown as (white) lines for comparison. Around the Fermi level there is a single Ni-3$d_{x^2-y^2}$ orbital, as in the one-band Hubbard model with an additional Nd-5$d_{xy}$ pocket around the $A$-point as a hardly hybridizing bystander. \label{Fig:DMFT}}
\end{figure*}

Hence we have two bands of predominately   Ni-3$d_{x^2-y^2}$ and   Nd-5$d_{xy}$ character. Their hybridisation is zero, see~Supplementary Information Table~S.III, which can be inferred already from the lack of any  splitting around the DFT crossing points in  Fig.~\ref{Fig:DMFT} (d,e).  There is some hybridisation of the  Nd-5$d_{xy}$  with  the other Ni-bands, which results in minor spectral weight of Nd-5$d_{xy}$ character in the region of the other Ni-3$d$ bands at -0.5 to -2.5\,eV in  Fig.~\ref{Fig:DMFT} (a-c) and vice-versa of the Ni-3$d_{z^2}$ orbital between 1 and 3\,eV. This admixing is however so minor, that it can be described by properly admixed, effective orbitals that are away from the Fermi energy, without multi-orbital physics.

If we study the doping dependence in Fig.~\ref{Fig:DMFT} (a-c), we see that with Sr-doping all bands move upwards. More involved and beyond a rigid-band picture,  also the   Ni-3$d_{x^2-y^2}$ band becomes less and less correlated. The effective mass enhancement  or inverse quasiparticle weight  changes from $m^*/m=1/Z=4.4$ for the undoped compound to  $m^*/m=2.8$ at 20\% Sr-doping  to   $m^*/m=2.5$  at 30\% Sr-doping. The Hubbard bands gradually disappear.

 At 30\% Sr-doping we are in the situation that the other Ni-3$d$ bands are now immediately below the Fermi energy. Hence around this doping it is no longer justified to employ a Ni-$d_{x^2-y^2}$-band plus  Nd-$d_{xy}$-pocket around $A$  picture to describe the low-energy physics. However below 30\% doping, the other Ni-3$d$ orbitals enter the stage. Because of the weak hybdization, the  Nd-$d_{xy}$-pocket only acts as an electron reservoir, which changes the doping of the   Ni-$d_{x^2-y^2}$-band from the lower $x$-axis in  Fig.~\ref{Fig:overview}\,(b) to the upper $x$-axis.

For this $d_{x^2-y^2}$-band  we have done a separate Wannier function projection
which results in the aforementioned hopping parameters $t=395\,$meV,  $t'=-95\,$meV, $t''=47\,$meV between next-nearest and the two next-next-nearest neighbours;  a cRPA calculation for this single orbital yields $U=3.2\,$eV$=8t$
\footnote{This value is slightly larger than the cRPA zero frequency
value $U=2.6\,$eV  \cite{Nomura2019,Hirofumi2019} to mimic the frequency
dependence as well as beyond cRPA contributions
\cite{Honerkamp2018}. In the Supplementary Information Section~S.5
we present further calculations within a realistic range of $U$ values.}. The hopping parameters in the $z$-direction are negligibly small $t_z=34\,$meV, leaving us with a --to a good approximation-- two-dimensional one-band Hubbard model.

This  two-dimensional one-band Hubbard model with the properly translated doping according to Fig.~\ref{Fig:overview}\,(b) can now be solved using more sophisticated, numerically expensive methods. We employ the D$\Gamma$A\cite{RMPVertex} in the following, which not only includes all local DMFT correlations, but also non-local correlations that are responsible for spin fluctuations\cite{RMPVertex} and $d$-wave superconductivity\cite{RMPVertex,Motoharu2019}. Please note that our D$\Gamma$A calculation also includes charge
    fluctuations \cite{Ghiringhelli2012} and their effect on
    superconductivity  on an equal footing to spin fluctuations,
    but the latter dominate. As a matter of course effects beyond the
    Hubbard model, such as disorder and phonons or a strengthening of
    charge fluctuations by non-local interactions which all are also
    considered to be of some relevance for superconductivity
    \cite{Keimer2015}  are not included. Studying a (tetragonal) rotational
    symmetry broken phase \cite{Kivelson1998}  requires a full parquet
    D$\Gamma$A or a eigenvalue analysis like we do for $d$-wave
    superconductivity here.

\begin{figure}[t]
\includegraphics[width=8.8cm]{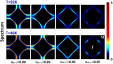}
\caption{D$\Gamma$A $\mathbf k$-resolved spectrum at the Fermi energy for $T=0.02t=92\,$K (upper panels) and  $T=0.01t=46\,$K (lower panels) and four different dopings $n_{d_{x^2-y^2}}$ of the Ni-$d_{x^2-y^2}$-band (left to right).\label{Fig:DGA}}
\end{figure}

Fig.~\ref{Fig:DGA} shows the thus obtained D$\Gamma$A Fermi surfaces at different dopings and two different temperatures. For the superconducting  Sr$_{0.2}$Nd$_{0.8}$NiO$_2$, which corresponds to $n_{d_{x^2-y^2}}=0.822$ electrons per site in the  Ni-$d_{x^2-y^2}$-band, we have a well defined hole-like Fermi surface, whereas for $n_{d_{x^2-y^2}}=0.9$ and, in particular for  $n_{d_{x^2-y^2}}=0.95$, we see the development of Fermi arcs induced by strong antiferromagnetic spin fluctuations.   The $A=(\pi,\pi,\pi)$-pocket is not visible in Fig.~\ref{Fig:DGA} because it is only included through the effective doping in the  D$\Gamma$A calculation. In any case it would be absent in the $k_z=0$ plane and only be visible around $k_z=\pi$. 
 
Whereas the Ni-3$d_{x^2-y^2}$-band is strongly correlated and has a Fermi surface that is prone to high-$T_{\rm C}$ superconductivity, the $A$-pocket is weakly correlated and hardly hybridises with the former. Nonetheless for some physical quantities, different from superconductivity it will play a role. For example, it will give  an electron-like negative contribution to the Hall coefficient.  
This will be partially compensated by the $d_{x^2-y^2}$-contribution which has a hole-like structure   for Sr$_{0.2}$Nd$_{0.8}$NiO$_2$, corresponding to $n_{d_{x^2-y^2}}=0.822$ which is in-between the two leftmost panels in Fig.~\ref{Fig:DGA}. Note that also for the cuprates a hole-like  Hall coefficient is found\cite{Putzke2019}. In contrast, for the undoped compound NdNiO$_2$ (i.e., $n_{d_{x^2-y^2}}=0.944\approx 0.95$), the  3$d_{x^2-y^2}$ Fermi surface in Fig.~\ref{Fig:DGA}  has a more electron-like shape but its Hall contribution should be suppressed because of the pseudo gap. 
This might explain why the Hall coefficient\cite{li2019superconductivity} is large and electron-like (negative) for   NdNiO$_2$, whereas it is smaller and changes from negative (electron-like) to positive (hole-like) below 50\,K. In  Fig.~\ref{Fig:DGA} we further see that for $T=92$\,K\,$=0.02t$, we have a strong scattering at the antinodal point $k=(\pi,0)$, whereas at the lower temperature   $T=46$\,K\,$=0.01t$, we have a well defined band throughout the Brillouin zone. This indicates that the hole-like contribution of the $d_{x^2-y^2}$-orbital becomes more important at lower temperatures, possibly explaining the experimental sign change of the Hall coefficient.

At the temperatures of Fig.~\ref{Fig:DGA},  Sr$_{x}$Nd$_{1-x}$NiO$_2$ is not yet superconducting. But we can determine $T_{\rm C}$ from the divergence of the superconducting susceptibility $\chi$, or alternative  the leading superconducting eigenvalue $\lambda_{\rm SC}$. These are related through, in matrix notation,   $\chi=\chi_0/[1-\Gamma_{pp}\chi_0]$. Here $\chi_0$ is the bare superconducting susceptibility and $\Gamma_{pp}$ the irreducible vertex in the particle-particle channel calculated by D$\Gamma$A; $\lambda_{\rm SC}$ is the leading eigenvalue of   $\Gamma_{pp}\chi_0$. If $\lambda_{\rm SC}$ approaches 1, the superconducting susceptibility is diverging.

\begin{figure}[t]
\includegraphics[width=7cm]{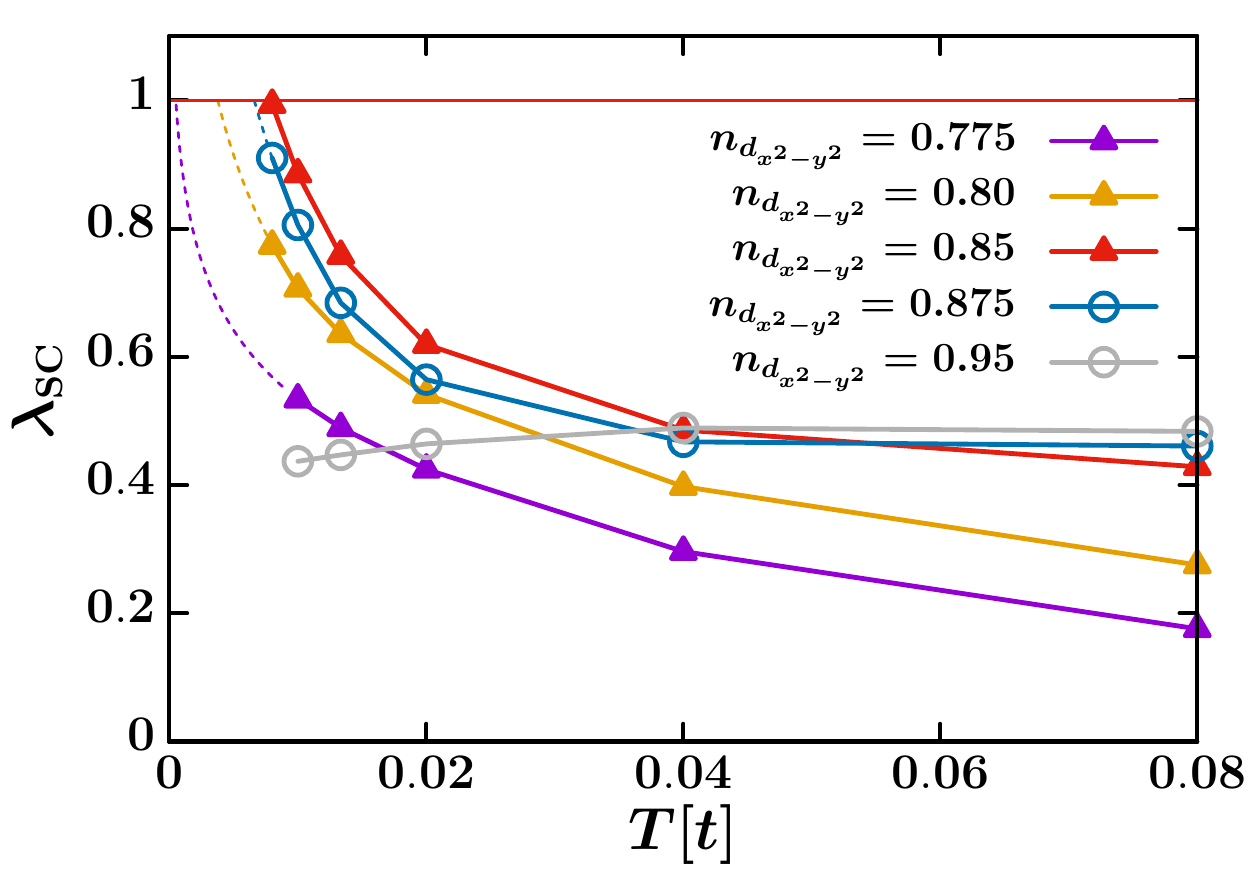}
\caption{Leading D$\Gamma$A superconducting eigenvalue $\lambda_{\rm SC}$ vs.~temperature $T$. The superconducting $T_{\rm C}$ corresponds to $\lambda_{\rm SC}\rightarrow 1$.\label{Fig:lambda}}
\end{figure}

In Fig.~\ref{Fig:lambda} we plot this  $\lambda_{\rm SC}$ vs.~temperature, and see that it approaches 1  at e.g.~$T=36K=0.008 t$ for $n_{d_{x^2-y^2}}=0.85$. But outside a narrow doping regime between $n_{d_{x^2-y^2}}=0.9$ and $0.8$ it does not approach 1;  there is no superconductivity.
Let us also note that the phase transition is toward $d$-wave superconductivity which can be inferred from the leading eigenvector corresponding to $\lambda_{\rm SC}$.  

Altogether, this leads to the superconducting dome of  Fig.~\ref{Fig:overview}\,(b). Most noteworthy Sr$_{0.2}$Nd$_{0.8}$NiO$_2$ which was found to be superconducting in experiment\cite{li2019superconductivity} is close to optimal doping  $n_{d_{x^2-y^2}}=0.85$ or Sr$_{0.16}$Nd$_{0.84}$NiO$_2$.
Our results call for a more thorough investigation of superconductivity in nickelates around this doping,
which is quite challenging experimentally\cite{Li2019,Zhou2019,Lee2020}.

Besides a slight increase of $T_{\rm C}$ by optimising the doping, and further room of improvement by adjusting $t'$ and $t''$, our results especially show that a larger bandwidth and a somewhat smaller interaction-to-bandwidth ratio may substantially enhance $T_{\rm C}$, see Supplementary Information  Fig.~S7. One way to achieve this is compressive strain, which enlarges the bandwidth while hardly affecting the interaction. Compressive strain can be realised by e.g.~growing thin nickelate films on a LaAlO$_3$ substrate with or without SrTiO$_3$ capping layer, or by Ca- instead of Sr-doping for the bulk or thick films. Another route is to substitute 3$d$ Ni by 4$d$ elements, e.g.\ in Nd(La)PdO$_2$ which has a similar Coulomb interaction and larger bandwidth~\cite{Botana2018,Motoaki2019}. 

 {\em Note added:} In an independent DFT+DMFT study Leonov {\em et
al.}\cite{Leonov2020} also observe the shift of the $\Gamma$ pocket
above $E_F$ with doping for LaNiO$_2$ and the occupation of further Ni 3$d$ orbitals besides the 3$d_{x^2-y^2}$ for large (e.g.~40\%) doping.

{\em Second note added:} Given that our phase diagram Fig.~\ref{Fig:overview}\,(b) has been a prediction with only a single experimental data point given, the most recently experimentally determined phase diagram  \cite{Li2020} turned out to be in very good agreement. In particular if one considers, as noted already above, that the theoretical calculation should overestimate $T_{\rm C}$, while the exerimentally observed  $T_{\rm C}$ is likely suppressed by extrinsic contributions such as disorder etc. Supplemental Information Section S.~6  shows a comparison, which corroborates  the modeling and theoretical understanding achieved in the present paper.

\begin{methods}
\subsection{Density functional theory (DFT)}
We mainly employ the  \textsc{wien2k} program package \cite{blaha2001wien2k} using the  PBE version of the generalised gradient approximation (GGA), a  $13\times13\times15$ momentum grid, and  $R_{MT}K_{max}$=7.0 with a muffin-tin radius $R_{MT}=$2.50, 1.95, and 1.68 a.u.\ for Nd, Ni, and O, respectively. But we also double checked against  \textsc{VASP}\cite{PhysRevB.48.13115}, which was also used for structural relaxation ($a=b=3.86\,$\AA{}, $c=3.24\,$\AA{})\cite{jiang2019electronic}, and \textsc{FPLO}\cite{FPLO}, which was used for Fig.~\ref{Fig:overview}~(a) and Supplementary Information Fig.~S.1.
The Nd-$4f$ orbitals are treated as open core states if not stated otherwise.
\subsection{Wannier function projection} 
The  \textsc{wien2k} bandstructure around the Fermi energy is  projected onto maximally localized Wannier functions\cite{Pizzi2020} using \textsc{wien2wannier} \cite{kunevs2010wien2wannier}. For the DMFT we employ a projection onto  five Ni-3$d$ and five Nd-5$d$ bands; for the parameterization of the one-band Hubbard model we project onto the Ni-3$d_{x^2-y^2}$ orbital. For calculating the hybridisation with the Nd-4$f$, we further Wannier projected onto a 17 bands, with the  Nd-4$f$ states now treated as valence bands in GGA.
\subsection{Dynamical mean-field theory (DMFT)} 
We supplement the  five Ni-3$d$ plus five Nd-5$d$ orbital  Wannier Hamiltonian by a cRPA calculated Coulomb repulsions \cite{Si2020} $U'= 3.10\,$eV ($2.00\,$eV) and Hund's exchange  $J=0.65\,$eV  (0.25\,eV) for Ni (Nd or La). The resulting 
Kanamori Hamiltonian is solved in  DMFT\cite{RevModPhys.68.13}  at room temperature (300\,K) using continuous-time quantum Monte Carlo simulations in the hybridisation expansions \cite{RevModPhys.83.349} implemented in \textsc{w2dynamics} \cite{wallerberger2019w2dynamics}.
 The maximum entropy method \cite{PhysRevB.44.6011} is employed for the analytic continuation of the spectra. 
\subsection{Dynamical vertex approximation} 
For calculating the superconducting $T_{\rm C}$, we employ the dynamical vertex approximation (D$\Gamma$A), for a review see Ref.~\onlinecite{RMPVertex}. We first calculate the particle-particle vertex with spin-fluctuations in the particle-hole and transversal-particle hole channel, and then the leading eigenvalues in the  particle-particle channel, as done before in Ref.~\onlinecite{Motoharu2019}. This is like the first iteration for the particle-particle channel in a more complete parquet D$\Gamma$A\cite{RMPVertex}.

\end{methods}




\begin{addendum}
 \item[] We thank 
A.~Hariki, J.~Kaufmann, F.~Lechermann, and J.~M.~Tomczak for helpful discussions; and U.~Nitzsche for technical assistance.  M.\,K. is supported by the RIKEN Special Postdoctoral Researchers Program; L.\,S.  and K.\,H.  by the Austrian Science Fund (FWF) through projects P\,30997 and P\,32044M; L.\,S.  by the China Postdoctoral Science Foundation (Grant No. 2019M662122);  L.\,S. and Z.\,Z. by the National Key R\&D Program of China (2017YFA0303602), 3315 Program of Ningbo, and the National Nature Science Foundation of China (11774360, 11904373); O.\,J. by the Leibniz Association through the Leibniz Competition.
R.\,A.  by Grant-in-Aid for Scientific Research (No. 16H06345, 19H05825) from MEXT, Japan.
 Calculations have been done on the Vienna Scientific Clusters (VSC). 
 \item[] The authors declare that they have no
competing financial interests.
\item [] $^*$ These authors contributed equally  to this work.
 \item[]  $^\dagger$ Correspondence and requests for materials
should be addressed to K.H.~(email: held@ifp.tuwien.ac.at).
\end{addendum}

\bibliography{full}

\end{document}


\title{Supplementary Information   ``Nickelate superconductors -- a renaissance of the one-band Hubbard model''}

\author{Motoharu Kitatani}
\affiliation{Institute for Solid State Physics, Vienna University of Technology, 1040 Vienna, Austria}
\affiliation{RIKEN Center for Emergent Matter Sciences (CEMS), Wako, Saitama, 351-0198, Japan}

\author{Liang Si}
\affiliation{Institute for Solid State Physics, Vienna University of Technology, 1040 Vienna, Austria}
\affiliation{Key Laboratory of Magnetic Materials and Devices \& Zhejiang Province Key Laboratory of Magnetic Materials and Application Technology, Ningbo Institute of Materials Technology and Engineering (NIMTE), Chinese Academy of Sciences, Ningbo 315201, China}

\author{Oleg Janson}
\affiliation{Leibniz Institute for Solid State and Materials Research IFW Dresden, 01171 Dresden, Germany}

\author{Ryotaro Arita}
\affiliation{RIKEN Center for Emergent Matter Sciences (CEMS), Wako, Saitama, 351-0198, Japan}
\affiliation{Department of Applied Physics, The University of Tokyo, Hongo, Tokyo, 113-8656, Japan}

\author{Zhicheng Zhong}
\affiliation{Key Laboratory of Magnetic Materials and Devices \& Zhejiang Province Key Laboratory of Magnetic Materials and Application Technology, Ningbo Institute of Materials Technology and Engineering (NIMTE), Chinese Academy of Sciences, Ningbo 315201, China}

\author{Karsten Held}
\affiliation{Institute for Solid State Physics, Vienna University of Technology, 1040 Vienna, Austria}

\date{\today}

\begin{abstract}
Here, we provide information supplementary to the main manuscript. Specifically, Section  \ref{Sec:orbcharact} presents 
the orbital character of the three bands crossing the Fermi level for undoped LaNiO$_2$ and NdNiO$_2$
in density functional theory.
Section~\ref{Sec:WF} discusses details of the Wannier function projections onto 1, 10 and  17 bands and tabulates the most important hopping parameters. 
Section~\ref{Sec:dopingDMFT} shows the dependence of the doping of the Ni-3$d_{x^2-y^2}$ band and the quasiparticle mass on Sr-doping as calculated in dynamical mean-field theory (DMFT).
Section~\ref{Sec:DMFTspectra} supplements the main text by additional DMFT spectra at other doping levels for NdNiO$_2$ as well as for  LaNiO$_2$.
Section~\ref{Sec:DGA} provides additional dynamical vertex approximation spectra and phase diagrams obtained for  a somewhat smaller and larger interaction $U$ than expected and employed in the main text. This hints that Nd(La)PdO$_2$ with a larger bandwidth and slightly smaller interaction might host even larger critical temperatures.
Section~\ref{Sec:comp} compares with the recently obtained experimental phase diagram.
\end{abstract}

\maketitle

\section{Orbital character of the important DFT bands}
\label{Sec:orbcharact}
Starting point of our analysis is the density functional theory (DFT) bandstructure which we have calculated by  \textsc{wien2k} \cite{blaha2001wien2k,schwarz02},  \textsc{VASP}\cite{PhysRevB.48.13115}, and \textsc{FPLO}\cite{FPLO} using the  PBE \cite{PhysRevLett.77.3865}  version of the generalized gradient approximation (GGA). There are three relevant orbitals that cross the Fermi energy $E_F=0$ in DFT. We determine their orbital character in \textsc{FPLO}\cite{FPLO}  version 18-00.55, using  a $k$-mesh of $19\times19\times22$ points for the scalar relativistic calculations. For NdNiO$_3$, the Nd-$4f$ states were put into the core (``open core''), their occupation is fixed to three electrons. At high-symmetry points, some orbitals may belong to the same two-dimensional irreducible representation. As a result, the respective orbital characters can show a seeming discontinuity in the band structure plot.

\begin{figure}[tb]
\includegraphics[width=18cm]{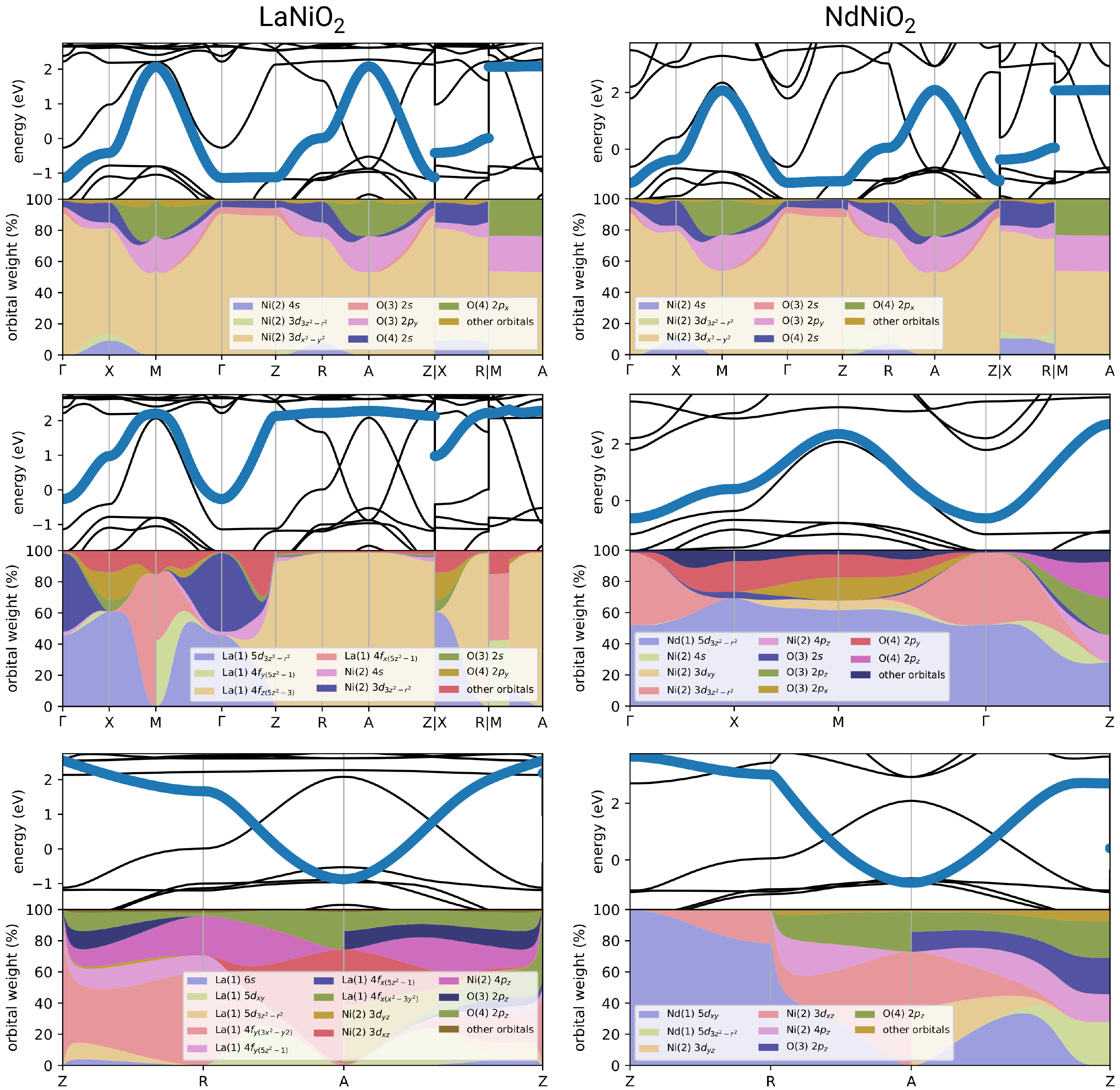}
\caption{Band weights for three bands that cross the Fermi energy $E_F=0$ for  LaNiO$_2$ (left panels) and NdNiO$_2$ (right panels), as calculated by DFT. In each panel, the top part shows the selected band (thick blue curve) along a path in the Brillouin zone, while the bottom part depicts the orbital composition (band characters) of the respective band along the same path. Here and in the following, the Fermi level is at $E_F=0$. 
\label{Fig:orbcharact}}
\end{figure}

The results are presented in  Fig.~\ref{Fig:orbcharact} for both, LaNiO$_2$ and
 NdNiO$_2$. The band in the first row is predominately of Ni-3$d_{x^2-y^2}$ character, with some oxygen character mixed in. This oxygen admixture is however much less than for the cuprates, where the oxygen orbitals are much closer, see Fig.~1 of the main paper. 
This band is close to half-filling and can be described, as we show in our paper, by  a one-band Hubbard model, if the doping is adjusted properly.
We casually  refer to it as the  ``Ni-3$d_{x^2-y^2}$'' band in the main text.

 The band in the second row forms a Fermi surface pocket around  the $\Gamma$-pocket. It is mainly of Nd(La)-5$d_{3z^2-r^2}$ character, but with much more admixture from the other orbitals, in particular, the Ni-3$d_{3z^2-r^2}$ orbital. In case of LaNiO$_2$ it is difficult to disentangle this band from the the La-4$f$ orbitals above 2\,eV, but this only concerns the less relevant upper band edge. In case of  NdNiO$_2$, these  Nd-4$f$ have been removed by treating them as part of the open core. 

An important observation is that the  $\Gamma$-pocket is only slightly below $E_F$ for 
 LaNiO$_2$ but extends noticeable below $E_F$ for NdNiO$_2$. This is however, possibly an artifact of the open core treatment. In a  plain GGA calculation
with Nd-4$f$ states as valance states (not shown), the distance between the bottom of the Nd-5$d_{3z^2-r^2}$ pocket and $E_F$ is more similar to that in LaNiO$_2$. In a DFT+$U$ treatment this pocket is spin-split due to a ferromagnetic interaction with the Nd-4$f$ magnetic moment \cite{Choi2020}.
The $\Gamma$-pocket disappears, i.e., is shifted above $E_F$, when including the  La-5$d$ interaction in DMFT for LaNiO$_2$\cite{Si2020}. Even  if we treat the  Nd-4$f$ as open core states, which stabilizes the $\Gamma$-pocket the most, the $\Gamma$ pocket disappears for Sr-doping above 10-15\% \footnote{When referring to doping, we have used the virtual crystal approximation.}.

Finally, there is the pocket around the 
$A$-point in  Fig.~\ref{Fig:orbcharact}, which extends much further below $E_F$ and hence remains present even up to a Sr-doping of 30\% in DMFT, see Fig.~\ref{Fig2} below.
 Hence it is this $A$-pocket which serves as an electron reservoir, with the important consequence that the doping of the  Ni-3$d_{x^2-y^2}$ orbital is different from the Sr-doping. This $A$-pocket can be associated in part of the Brillouin zone with the  Nd-5$d_{xy}$ orbital. However it is quite intermixed with other orbitals and at its bottom (around the $A$-point) it crosses the  Nd-3$d$ orbitals which makes it difficult to trace.

In the next Section, we will see that the hybridization with the   Ni-3$d_{x^2-y^2}$ orbital is vanishing so that the $A$-pocket can be considered as an electron reservoir which otherwise does not affect the strongly correlated   Ni-3$d_{3z^2-r^2}$ orbital (or the one-band Hubbard model description).
The proper translations between the Sr-doping  and that of the  Ni-3$d_{x^2-y^2}$ orbital is given in Fig.~1 (right) of the main text and in Fig.~\ref{Fig:doping} below.

\section{Wannier function projections}
\label{Sec:WF}

\begin{figure*}[h]
\includegraphics[width=10cm]{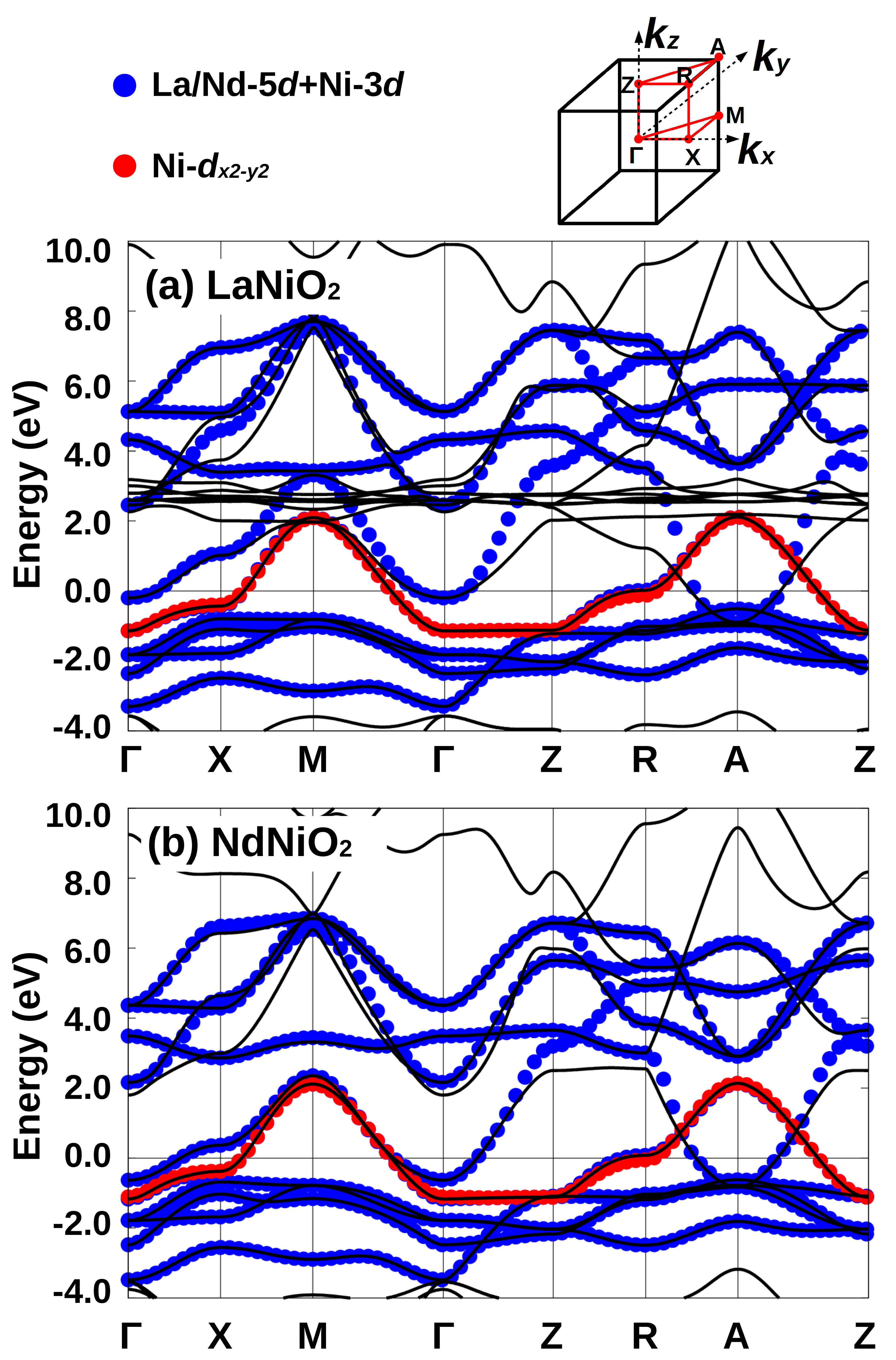}
\caption{Bandstructure of LaNiO$_2$ (a) and NdNiO$_2$ (b). For NdNiO$_2$, we treat the  Nd-4$f$ orbitals as open core states. Single-band (Ni-$d_{x^2-y^2}$, red dots) and 10-bands (La/Nd-5$d$+Ni-3$d$, blue dots) Wannier bands are superimposed on the DFT bandstructure (black lines). } 
\label{Fig4}
\end{figure*}

Next, we present details of the Wannier function projection, which has been done   using \textsc{wien2wannier} \cite{mostofi2008wannier90,kunevs2010wien2wannier} for the projection of the  \textsc{wien2k} bandstructure onto
maximally localized  Wannier functions \cite{PhysRev.52.191,RevModPhys.84.1419,Pizzi2020}.
Fig.~\ref{Fig4} provides for an overview, showing the DFT bandstructure together with the Wannier function projection on 10-bands (La/Nd-5$d$+Ni-3$d$; blue dots) and on  the  Ni-$d_{x^2-y^2}$ band only (red dots). The former projection is used for the subsequent dynamical mean-field theory (DMFT) multi-orbital calculations. These calculations however show that a properly doped Ni-$d_{x^2-y^2}$-band  (one-band Hubbard model) description is sufficient. Hence, we have also performed a projection onto this one-band only.  We have further done Wannier projections onto 17 orbitals  (Nd-4$f$+Nd-5$d$+Ni-3$d$) which is not included Fig.~\ref{Fig4}, as it starts from the plain GGA calculation which puts the Nd-4$f$ states just above $E_F$. For this GGA calculation, the Wannier projections onto 1- and 10-band have been done as well.

In Table~\ref{table1}, we present the hopping parameters of the Ni-3$d_{x^2-y^2}$ orbital which forms the one-band Hubbard model, as obtained for LaNiO$_2$, NdNiO$_2$ as well as for LaPdO$_2$, treating the Nd-$4f$ either as valence bands in GGA or as open core in GGA. Here $t_{R_x,R_y,R_z}$ denotes the hopping by $R_i$ unit cells in the $i$ direction. That is, $t_{000}$ is the on-site potential, $t=-t_{100}$ the nearest neighbor hopping, $t'=-t_{110}$ and $t''=-t_{200}$ the next nearest neighbor hopping, and $t_z=-t_{001}$ the hopping in the $z$-direction perpendicular to the NiO$_2$ plane. The hopping parameters are surprisingly similar for LaNiO$_2$ and  NdNiO$_2$ and the different Wannier projections, considering the fact that e.g.~the 4$f$ orbitals are at very different energies for the three DFT calculations in Table~\ref{table1}.   LaPdO$_2$, on the other hand, with 4$d_{x^2-y^2}$  instead of  3$d_{x^2-y^2}$ has a much larger bandwidth.
 In the main text, we give the values for the open core GGA 1-band Wannier projection for  NdNiO$_2$.

\begin{table}[tb]
\begin{tabular}{c|c|c|c|c|c|c}
\hline
\hline
LaNiO$_2$ (GGA) & $t_{000}$&$t_{100}$&$t_{001}$&$t_{110}$&$t_{200}$&$t_{210}$  \\
\hline
1-band (Ni-$d_{x^2-y^2}$) &  0.2689 & -0.3894 & -0.0362 & 0.0977 & -0.0465 & -0.0037 \\

10-bands (La-$d$+Ni-$d$)   &  0.2955 & -0.3975 & -0.0458 & 0.0985 & -0.0491 & 0.0000 \\

17-bands (La-$f$+La-$d$+Ni-$d$)   &   0.3514  &  -0.3943 & -0.0239 & 0.0792  & -0.0422  & -0.0008  \\
\hline
\hline
NdNiO$_2$ (GGA) & $t_{000}$&$t_{100}$&$t_{001}$&$t_{110}$&$t_{200}$&$t_{210}$  \\
\hline
1-band (Ni-$d_{x^2-y^2}$) &  0.2502 & -0.3974 & -0.0287 & 0.0933 & -0.0474 & -0.0027 \\

10-bands (Nd-$d$+Ni-$d$)   &   0.1998 & -0.4068  & -0.0763  &  0.1007 & -0.0428  &  0.0015 \\

17-bands (Nd-$f$+Nd-$d$+Ni-$d$)   &   0.2982  &  -0.4065 &  -0.0289 &  0.0773 & -0.0429  & 0.0026  \\
\hline
\hline
NdNiO$_2$ (GGA open core) & $t_{000}$&$t_{100}$&$t_{001}$&$t_{110}$&$t_{200}$&$t_{210}$  \\
\hline
1-band (Ni-$d_{x^2-y^2}$) &   0.3058  & -0.3945  &  -0.0336 & 0.0953  & -0.0471 &  -0.0031 \\

10-bands (Nd-$d$+Ni-$d$)   &  0.3168   &  -0.3976 & -0.0389  &  0.0949  &  -0.0480  & -0.0008   \\
\hline
\hline
LaPdO$_2$ (GGA) & $t_{000}$&$t_{100}$&$t_{001}$&$t_{110}$&$t_{200}$&$t_{210}$  \\
\hline
1-band (Ni-$d_{x^2-y^2}$) &   0.4094  & -0.5373  &  -0.0448 & 0.0975 &  -0.0708 &  -0.0058\\
\hline
\hline
\end{tabular}
\caption{Major hopping elements (in units of eV)  of the Ni-3$d_{x^2-y^2}$ orbital from 1-band (Ni-3$d_{x^2-y^2}$), 10-bands (La/Nd-$d$+Ni-$d$) and 17-bands (La/Nd-$f$+La/Nd-$d$+Ni-$d$) model projections. In the last two lines, we also show the hopping parameters for LaPdO$_2$. The 
DFT-relaxed lattice parameters are: LaNiO$_2$  ($a=b=3.88\,$\AA{}, $c=3.35\,$\AA{}),  NdNiO$_2$  ($a=b=3.86\,$\AA{}, $c=3.24\,$\AA{}),  LaPdO$_2$  ($a=b=4.13\,$\AA{}, $c=3.27\,$\AA{}).
\label{table1}}
\vspace{.6em}

\begin{tabular}{c|c|c|c|c|c|c|c}
\hline
\hline
LaNiO$_2$ (GGA) & $f_{xz^2}$ & $f_{yz^2}$ & $f_{z^3}$ & $f_{x(x^2-3y^2)}$ & $f_{y(3x^2-y^2)}$ & $f_{z(x^2-y^2)}$ & $f_{xyz}$ \\
\hline
Ni-$d_{x^2-y^2}$ & -0.0300   & 0.0300 & 0.0000 & -0.0851 & -0.0851 & -0.0203 & -0.0000 \\
\hline
\hline
NdNiO$_2$ (GGA) & $f_{xz^2}$ & $f_{yz^2}$ & $f_{z^3}$ & $f_{x(x^2-3y^2)}$ & $f_{y(3x^2-y^2)}$ & $f_{z(x^2-y^2)}$ & $f_{xyz}$ \\
\hline
Ni-$d_{x^2-y^2}$ & -0.0215  & 0.0215 & 0.0000 & -0.0612 & -0.0612 & 0.0160 & -0.0000   \\
\hline
\hline
\end{tabular}
\caption{Hybridization (hopping amplitude in eV) between the Ni-3$d_{x^2-y^2}$ and the Nd(La)-4$f$ orbitals, as  obtained from Wannier projections onto 17-bands (La/Nd-4$f$+La/Nd-5$d$+Ni-3$d$) including the  4$f$ as valence states in DFT(GGA).
\label{table2}}
\vspace{.6em}

\begin{tabular}{c|c|c|c|c|c}
\hline
\hline
LaNiO$_2$  & La-$d_{xy}$ & La-$d_{yz}$ & La-$d_{xz}$ & La-$d_{x^2-y^2}$ & La-$d_{z^2}$  \\
\hline

Ni-$d_{x^2-y^2}$ (10-bands model, GGA) & 0.0000  &  0.0835 & -0.0835 & -0.0168  & 0.0000  \\

Ni-$d_{x^2-y^2}$ (17-bands model, GGA) & 0.0000  &  0.0846  & -0.0846  & -0.0372 & 0.0000  \\

\hline
\hline
NdNiO$_2$ & Nd-$d_{xy}$ & Nd-$d_{yz}$ & Nd-$d_{xz}$ & Nd-$d_{x^2-y^2}$ & Nd-$d_{z^2}$  \\
\hline

Ni-$d_{x^2-y^2}$ (10-bands model, GGA with open core) & 0.0000  &   0.0701 & -0.0701 & -0.0388   & 0.0000  \\

Ni-$d_{x^2-y^2}$ (10-bands model, GGA) & 0.0000  &   0.0775 & -0.0775  & -0.0066    & 0.0000  \\

Ni-$d_{x^2-y^2}$ (17-bands model, GGA) & 0.0000  & 0.0811   & -0.0811  & -0.0239 & 0.0000  \\
\hline
\hline
\end{tabular}
\caption{Hybridization (hopping amplitude in eV) between the  Ni-$d_{x^2-y^2}$ and the La/Nd-5$d$ orbitals. The results are obtained from Wannier projections onto 17-bands (La/Nd-4$f$+La/Nd-5$d$+Ni-3$d$) and 10-bands  (La/Nd-5$d$+Ni-3$d$). 
The  Nd-4$f$ bands are treated  as core states in ``GGA with opencore'' and as valence states in ``GGA''.
\label{table3}}
\end{table}

A further relevant result of the Wannier projection is the hybridization with the 4$f$ orbitals which is shown in Table~\ref{table2}. These hybridizations are rather small, maximally $V=60$\,meV for NdNiO$_2$. Such a hybridization is by far too small to give rise to a Kondo effect. Even if we take this maximal hybridization and a typical 4$f$-Coulomb interaction of $U=5\,eV$, the Kondo coupling is only $J=4V^2/U=3\,meV$. Which even,  yields an exponential factor\cite{Hewson1993} of only $e^{-1/(N\rho_0*J)}\approx  10^{-54}$ for the Kondo temperature (taking a typical $\rho_0=0.2$\,eV$^{-1}$ from Fig. 2 of the main text and $N=2\times 7$ as a maximal upper bound). The prefactor is of the order of 1\,eV or smaller, so that we can conclude  that there is no Kondo effect between the localized Nd-4$f$ moments and the Ni 3$d_{x^ 2-y^ 2}$ orbital.

Next we turn to the hybridization between the relevant Ni-3$d_{x^2-y^2}$ band and the  La/Nd-5$d$ orbitals. The important observation is that the hybridization between the  Ni-3$d_{x^2-y^2}$ band and the Nd(La)-5$d_{xy}$ and the Nd(La)-5$d_{z^2}$ is zero.
These are the most important hybridizations since these two Nd(La) orbitals  form the basis of the  $A$- and $\Gamma$-pocket, respectively. 
Hence we can, to a very good approximation, indeed consider these pockets to be independent of the Ni-3$d_{x^2-y^2}$ band, except for that they may serve as an electron reservoir. Part of the holes, induced by e.g.~Sr-doping will go into the  Ni-3$d_{x^2-y^2}$ band, and part into the $A$- and $\Gamma$-pocket. The latter is completely depopulated in the superconducting regime.

\section{Doping and mass enhancement of the Ni-$d_{x^2-y^2}$ orbital in DMFT}
\label{Sec:dopingDMFT}
\begin{figure*}[h]
\includegraphics[width=15cm]{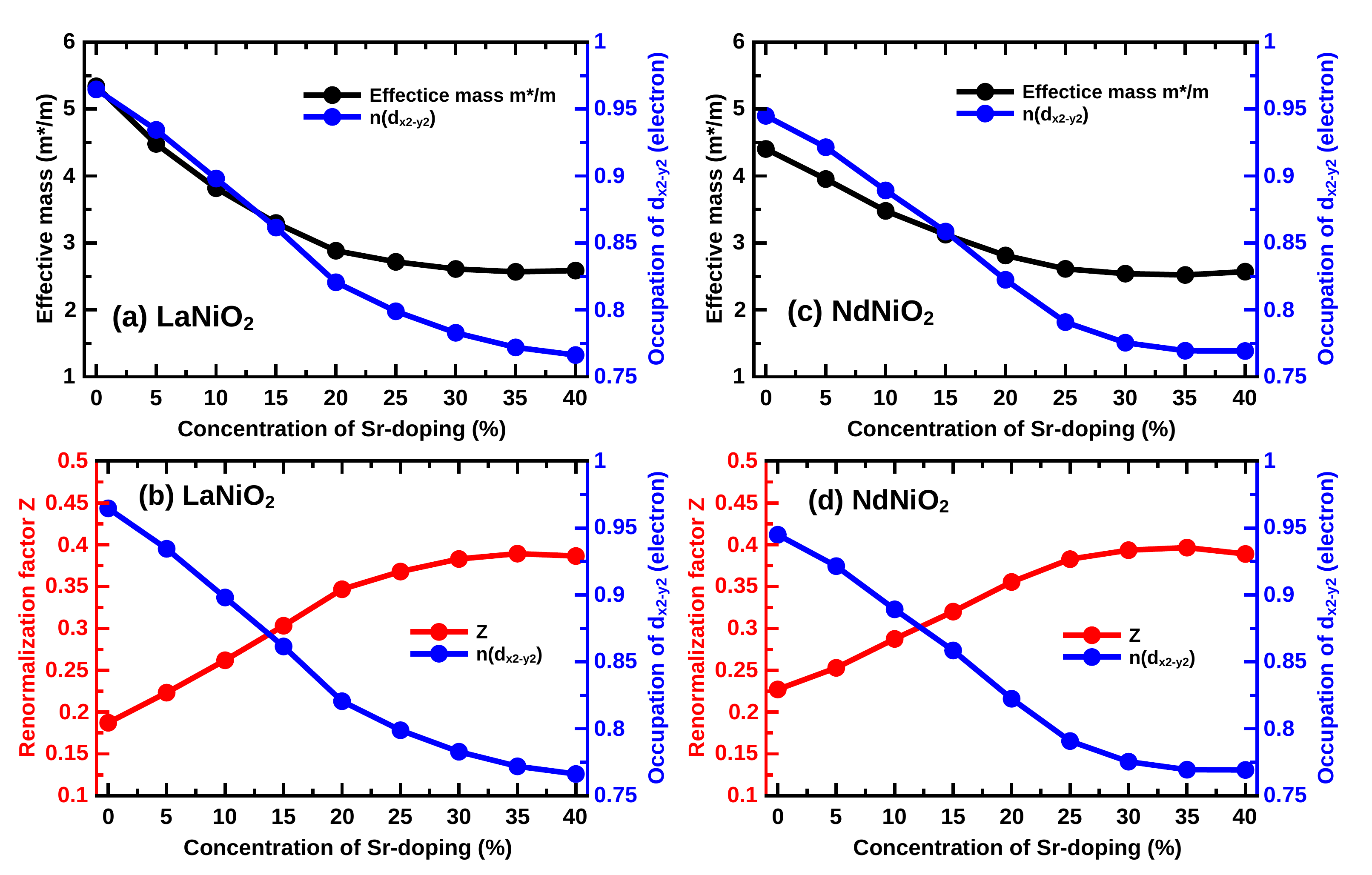}
\caption{ Occupation of the Ni-$d_{x^2-y^2}$ orbital [blue; right $y$-axis], its  effective mass enhancement m*/m  [black ; left $y$-axis in panels (a,c)] and quasiparticle  renormalization  $Z$  [red ; left $y$-axis in panels (b,d)] vs.~Sr-doping
for LaNiO$_2$ (left) and NdNiO$_2$ (right).}
\label{Fig1}\label{Fig:doping}
\end{figure*}

In Fig.~1 of the main text, the scales of the lower and upper $x$-axis already provide a translation between the Sr-doping (in the virtual crystal approximation) and the doping of the Ni-$d_{x^2-y^2}$ orbital (or the one-band Hubbard model). In Fig.~\ref{Fig1} we additionally show the functional dependence explicitly, and also compare the NdNiO$_2$ with the LaNiO$_2$ compound. For both compounds we see that a  similar amount of $\sim60\%$ of the holes go to the  Ni-$d_{x^2-y^2}$ orbital. For (very) small dopings a little bit less so in case for NdNiO$_2$ than for LaNiO$_2$ because we also have to depopulate the $\Gamma$-pocket, whereas this is already shifted above $E_F$ for  LaNiO$_2$ if we include the La-5$d$ interaction in DMFT\cite{Si2020}. The mass enhancement gradually decreases with the doping of the $d_{x^2-y^2}$ orbital, as close to half-filling electronic correlations are strongest. For the parent compound  LaNiO$_2$ the mass enhancement is slightly larger than for NdNiO$_2$, which agrees with the observation that it is closer to half-filling.

\section{Additional DMFT spectra}
\label{Sec:DMFTspectra}
\begin{figure*}[h]
\includegraphics[width=18cm]{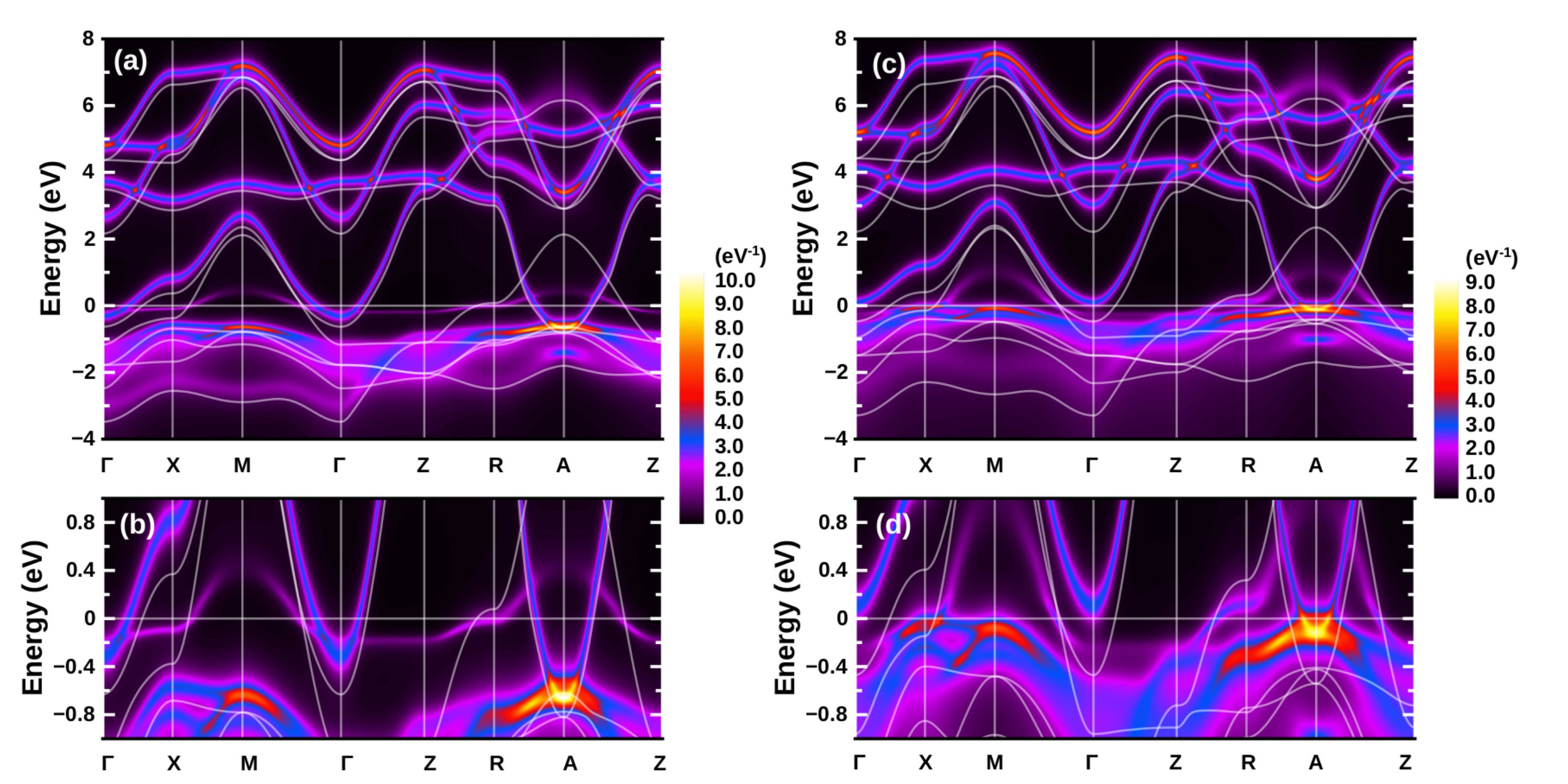}
\caption{DMFT $k$-revolved spectral function $A(k,\omega)$ of undoped NdNiO$_2$ (a-b) and 30\% Sr-doped NdNiO$_2$, i.e., Sr$_{0.3}$Nd$_{0.7}$NiO$_2$ (c-b). Panels (b,d) are zoom-ins of (a,c). White lines are the corresponding Wannier bands.} 
\label{Fig2}
\end{figure*}

In the main text, we have already shown the $k$-dependent DMFT spectral function of Sr$_{0.2}$ Nd$_{0.8}$NiO$_2$ in Fig.~2. In
Fig.~\ref{Fig2} we supplement this with the spectral functions at 0 and 30\% Sr-doping. For those Sr-dopings we have already presented the $k$-integrated spectra in  Fig.~2 of the main text. One sees that for the parent compound, NdNiO$_2$, there is a $\Gamma$ pocket whereas it is shifted above $E_F$ for Sr$_{0.3}$Nd$_{0.7}$NiO$_2$, as well as for  Sr$_{0.2}$Nd$_{0.8}$NiO$_2$ in the main text.
For   Nd$_{0.7}$Sr$_{0.3}$NiO$_2$ the other Ni-3$d$ bands almost touch $E_F$ at the $A$-point. Hence around this doping a one-band Hubbard model description is not possible any longer, all Ni orbitals and the interaction among  these needs to be taken into account.
 
\begin{figure*}[h]
\includegraphics[width=18cm]{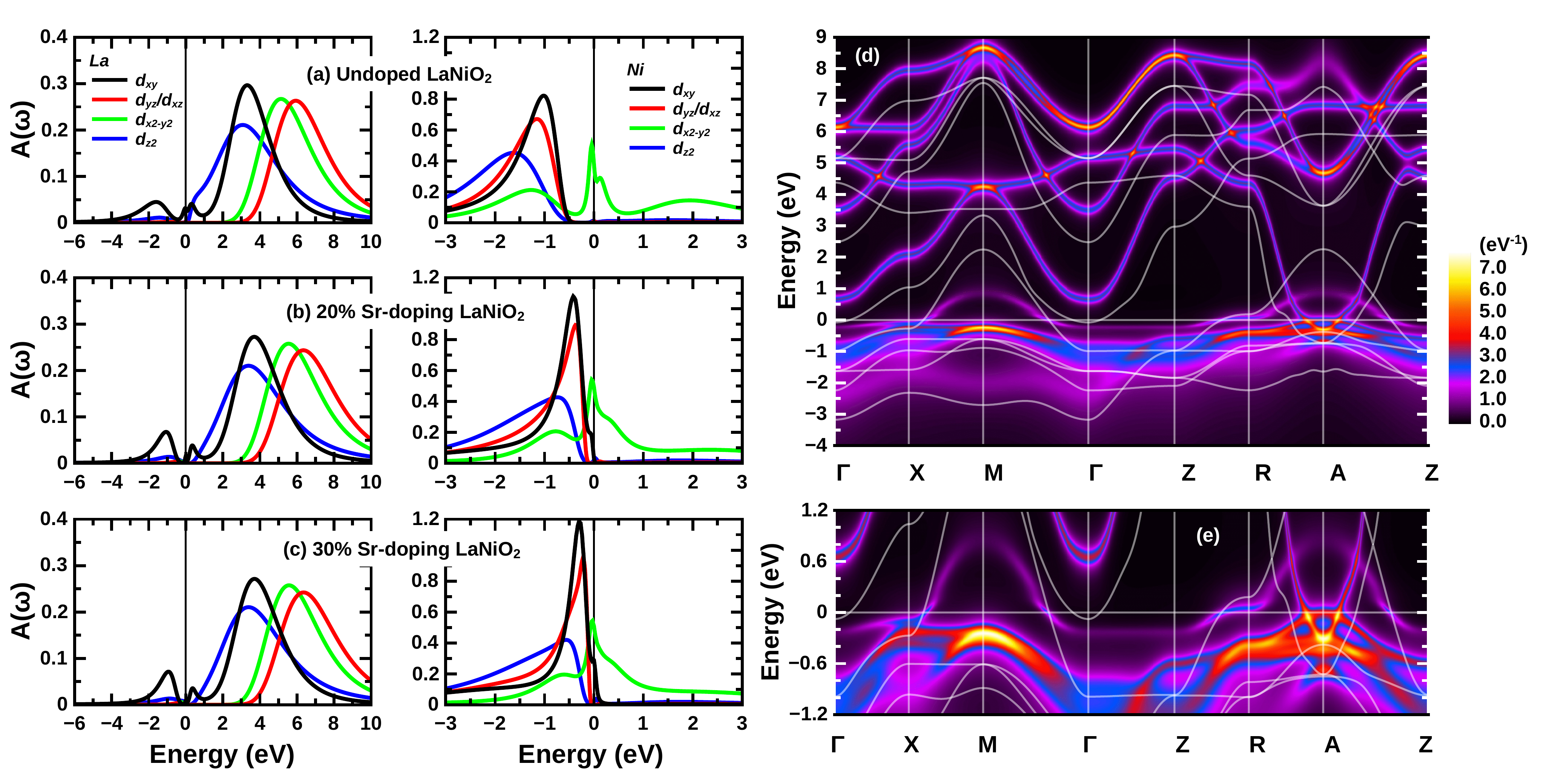}
\caption{DMFT $k$-integrated (a-c) and $k$-revolved (d-e) spectral functions $A(\omega)$ and $A(k,\omega)$ of undoped LaNiO$_2$ (a), 20\%Sr-doped LaNiO$_2$ (La$_{0.8}$Sr$_{0.2}$NiO$_2$) (b) and 30\%Sr-doped LaNiO$_2$ (La$_{0.7}$Sr$_{0.3}$NiO$_2$) (c). 
The  $k$-resolved spectral function $A(k,\omega)$ of La$_{0.8}$Sr$_{0.2}$NiO$_2$ is shown in (d); (e) is a  zoom-in of (d). } 
\label{Fig3}
\end{figure*}

With  Fig.~\ref{Fig3}, we provide for exactly the same overview of the DMFT results as in Fig.~2 of the main text but now for LaNiO$_2$ instead of NdNiO$_2$. An important difference is that for LaNiO$_2$ the Ni-$d_{xy}$ is immediately below $E_F$  already at 20\% Sr-doping.
At 30\% Sr-doping it accommodates already many holes and crosses the Fermi energy. Hence, in case of Sr-doped LaNiO$_2$ the one-band Hubbard model description is only valid up to about Sr-20\% doping. Another difference is that the $\Gamma$ pocket, which was present in DFT, is  shifted considerably  higher up in energy  in Fig.~\ref{Fig3}~(d,e). This agrees with the aforementioned obervation that it is absent already for the parent compound LaNiO$_2$.

\section{$U$-dependence of the D$\Gamma$A results}
\label{Sec:DGA}
In this section, we show the dependence of our  D$\Gamma$A results within a reasonable range of $U$ values above the cRPA value\cite{Nomura2019,Hirofumi2019} of $U=2.6\,$eV$=6.7t$. Specifically, in addition to $U=8t$ (in the main text), we consider
$U=7t$ and $9t$. As already pointed out in the main manuscript, the adequate value should be a bit larger than the cRPA value if we disregard the frequency dependence of $U$. These $U=7t$ and $U=9t$ values are the smallest and largest $U$ value,  respectively, which we consider still conceivable given  the cRPA calculated value. As for the hopping parameters we have employed the rounded ratios $t'/t=-0.25$ and $t''/t=0.12$ in D$\Gamma$A. 

\begin{figure}
\begin{centering}
\includegraphics[width=1.0\columnwidth]{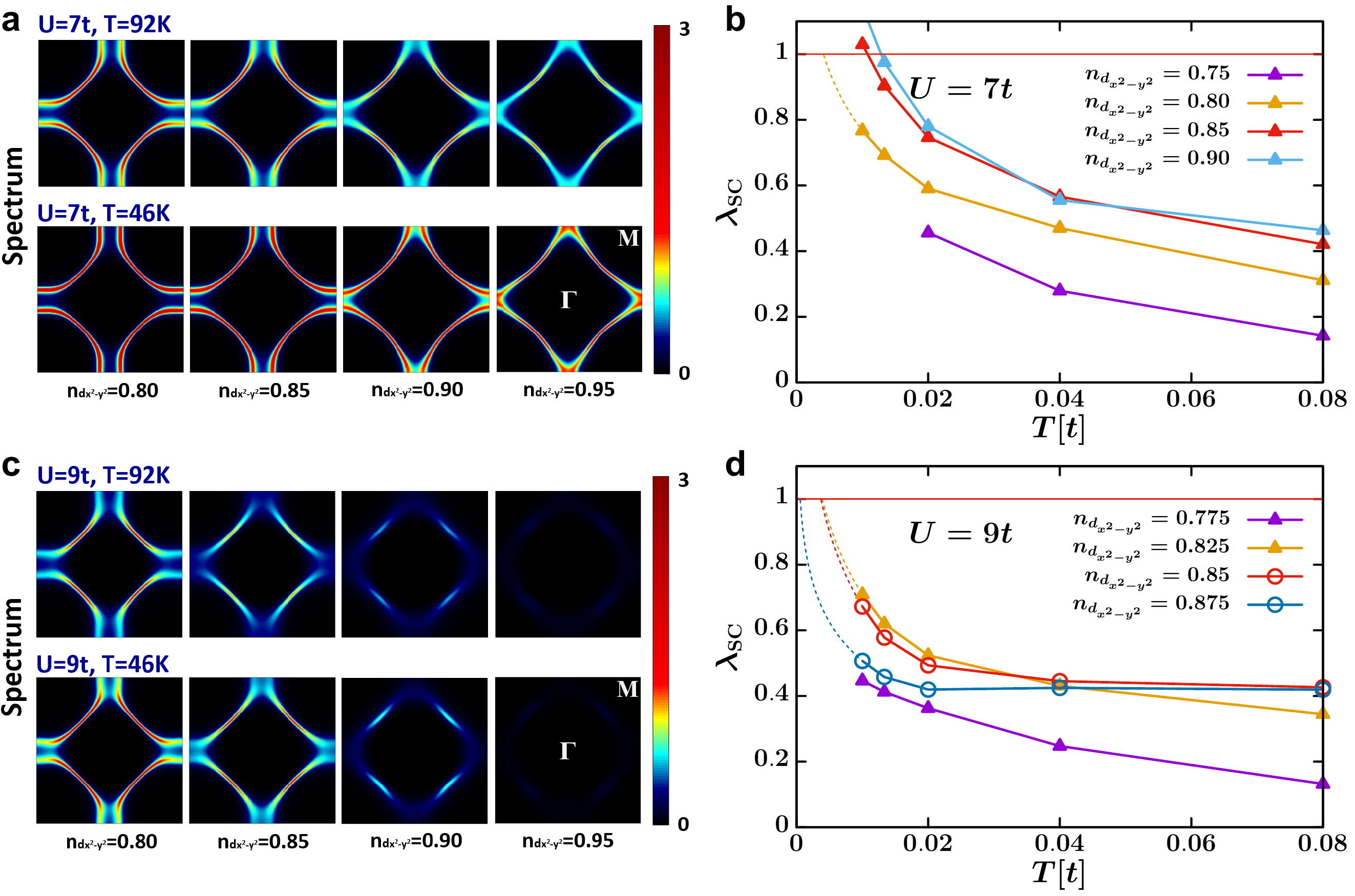}
\par\end{centering}
\caption{(a) Imaginary part of the Green function at the lowest Matsubara frequency $A(\mathbf{k},\omega_0 \equiv \pi/\beta) \equiv-\Im G(\mathbf{k},\omega_0)/\beta$ and (b) temperature dependence of the superconducting eigenvalue ($\lambda_{\rm SC}$) for $U=7t$. (c,d) Same figures but for $U=9t$.}
\label{fig:U7-9}
\end{figure}

In Fig.~\ref{fig:U7-9}, we show the momentum dependence of the $\mathbf{k}$-resolved  spectrum and the superconducting eigenvalue ($\lambda_{\rm SC}$) vs. temperature $(T)$, which are the same plot as the Fig.~3 in the main text, but  now at $U=7t$ and $U=9t$ instead of  $U=8t$. 
For $U=7t$, the self-energy damping effect becomes smaller and we can still see the Fermi surface (peak of the spectrum) around $\mathbf{k}=(0,\pi)$ even for low dopings. The superconducting eigenvalue $\lambda_{\rm SC}$ and hence the superconducting susceptibility is slightly  increasing towards low doping: $n_{d_{x^2-y^2}}=0.90$. For $U=9t$, we see that the damping effect becomes stronger instead and there is  a strong momentum dependence even for large doping ($n_{d_{x^2-y^2}}=0.80$). A consequence of this increased damping is that  $\lambda_{\rm SC}$ becomes smaller than for $U=7$ and $U=8t$. 

We also show the phase diagram for these three $U$ values in Fig.~\ref{fig:suppl-phase}. As discussed in the main paper the phase diagrams are obtained from  $\lambda_{\rm SC}\rightarrow 1$, which for higher $T_{\rm C}$'s is interpolated and for lower $T_{\rm C}$'s extrapolated with a fit function\cite{Sekine2013} of the form $a-b\ln(T)$. This fit function is plotted as a dashed line (also in Fig.~4 of the main text). As a matter of course the extrapolation over a large temperature interval, i.e. for  n$_{d_{x^2-y^2}}=0.775$ in Fig.~4 of the main text and  n$_{d_{x^2-y^2}}=0.875$ in ~\ref{fig:suppl-phase} (d), leads to a large uncertainty.  We hence cannot say whether $T_{\rm C}$  is still finite for n$_{d_{x^2-y^2}}=0.775$  or whether we are already outside the superconducting regime at this doping.

The physical conclusions from Fig.~\ref{fig:suppl-phase} are: $T_{\rm C}$ goes down and the optimal doping level moves to the larger doping side as we go toward stronger interaction. At 20\% Sr-doping, all results indicate a bit higher $T_{\rm C}$ than the experimental $T_{\rm C}$.
The result for $U=9t$ would be very close to the experimental result, but we think that 
the theoretical $T_{\rm C}$ has to be larger than the experimental one  (for the reasons stated in the main manuscript). This also agrees with our expectation that $U=8t$ is probably the best estimate for a static  interaction parameter.
Our results indicate that nickelates are in  the strong-coupling, larger $U$ regime with a dome-shaped $T_{\rm C}$ vs.~$U$ phase diagram, similar  as is considered to be the situation for the cuprates.

An important conclusion from Fig.~\ref{fig:suppl-phase} is that larger critical $T_{\rm C}$'s can be obtained by enhancing the bandwidth and reducing the ratio of Coulomb interaction to bandwidth. This can be achieved by strain or by replacing  Nd(La)NiO$_2$ with  Nd(La)PdO$_2$ which has $t=-t_{100}= 537\,$meV instead of $t=395\,$meV [see Table~\ref{table1}]. The cRPA $U$ value does not change that strongly\footnote{See Ref.~\onlinecite{Motoaki2019} which considers two Pd-compounds with 4d$^9$ configuration. These have a somewhat smaller and larger $U$ value than NdNiO$_2$. Note that the bandwidth of NdPdO$_2$ is even larger than for the compounds considered in  Ref.~\onlinecite{Motoaki2019} .}, so that
we have $U/t\approx 6$ for  Nd(La)PdO$_2$  instead of  $U/t\approx 8$ for  Nd(La)NiO$_2$.

\section{Comparison with experimental phase diagram}
\label{Sec:comp}
{\em Added:} Here, we plot on top of  Fig.~1(b) of the main text also the recently determined experimental phase diagram of Sr$_x$Nd$_{1-x}$NdO$_2$ by  Li {\em et al.}\cite{Li2020}. Since we calculate the onset of the second order phase transition to the superconducting phase, we compare to the experimental onset of the phase transition ($T_{{\rm c}, 90\%R}$ in Ref.\onlinecite{Li2020}) and the upper limit in Ref.~\onlinecite{li2019superconductivity}. As already mentioned in the last section, the extrapolation for $n_{d_{x^2-y^2}}= 0.775$ (22.5\% doping of the Hubbard model, $\approx 30\%$ Sr dopign)  covers a too large temperature range to say for sure whether $T_{\rm C}=0$ or small but finite.

\begin{figure}[t]
\begin{centering}
\includegraphics[width=0.4\columnwidth]{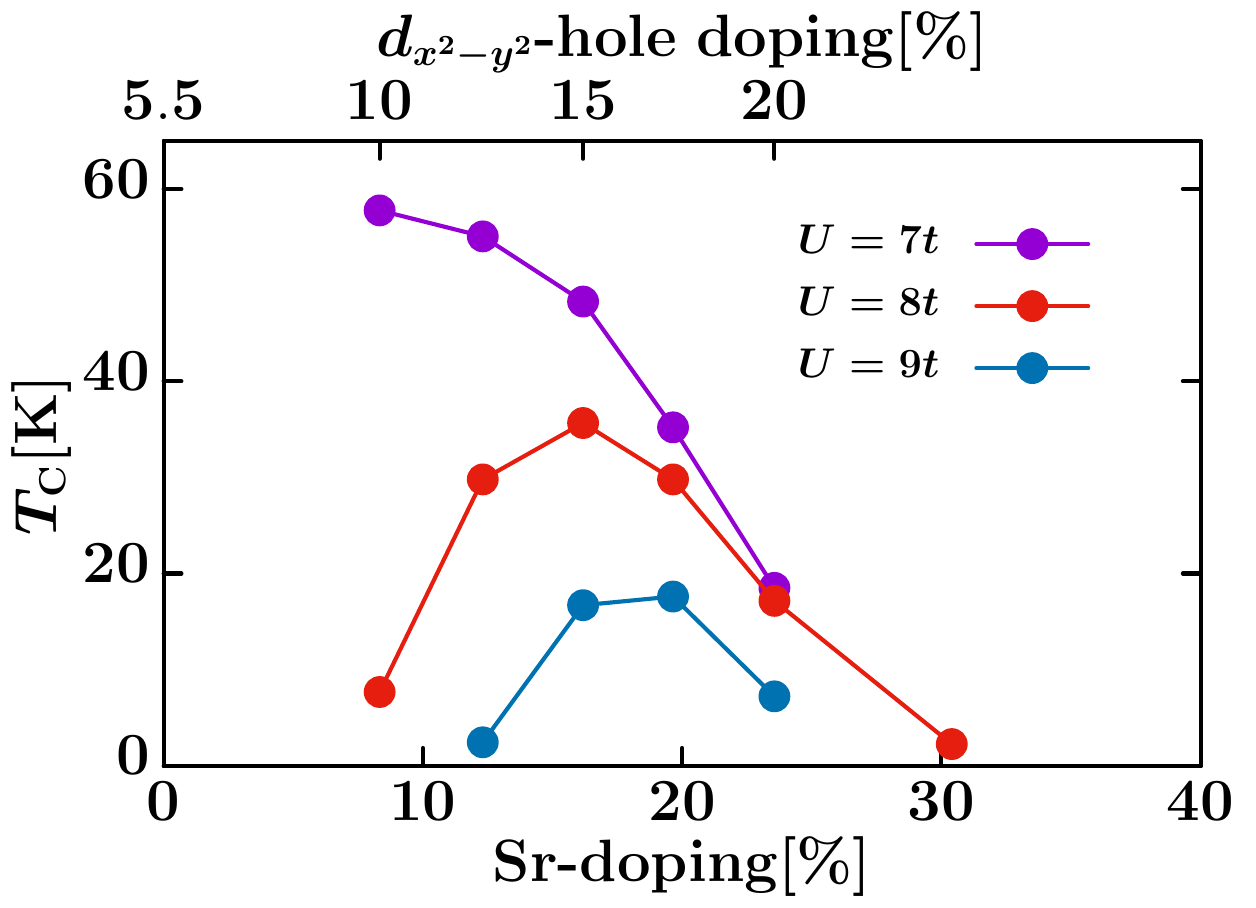}
\par\end{centering}
\caption{Superconducting $T_{\rm C}$ vs.~Sr-doping ($d_{x^2-y^2}$ filling on the upper $x$-axis) comparing $U= 7t, 8t$ and $9t$.}
\label{fig:suppl-phase}
\end{figure}

\begin{figure}[t]
\begin{centering}
\includegraphics[width=0.4\columnwidth]{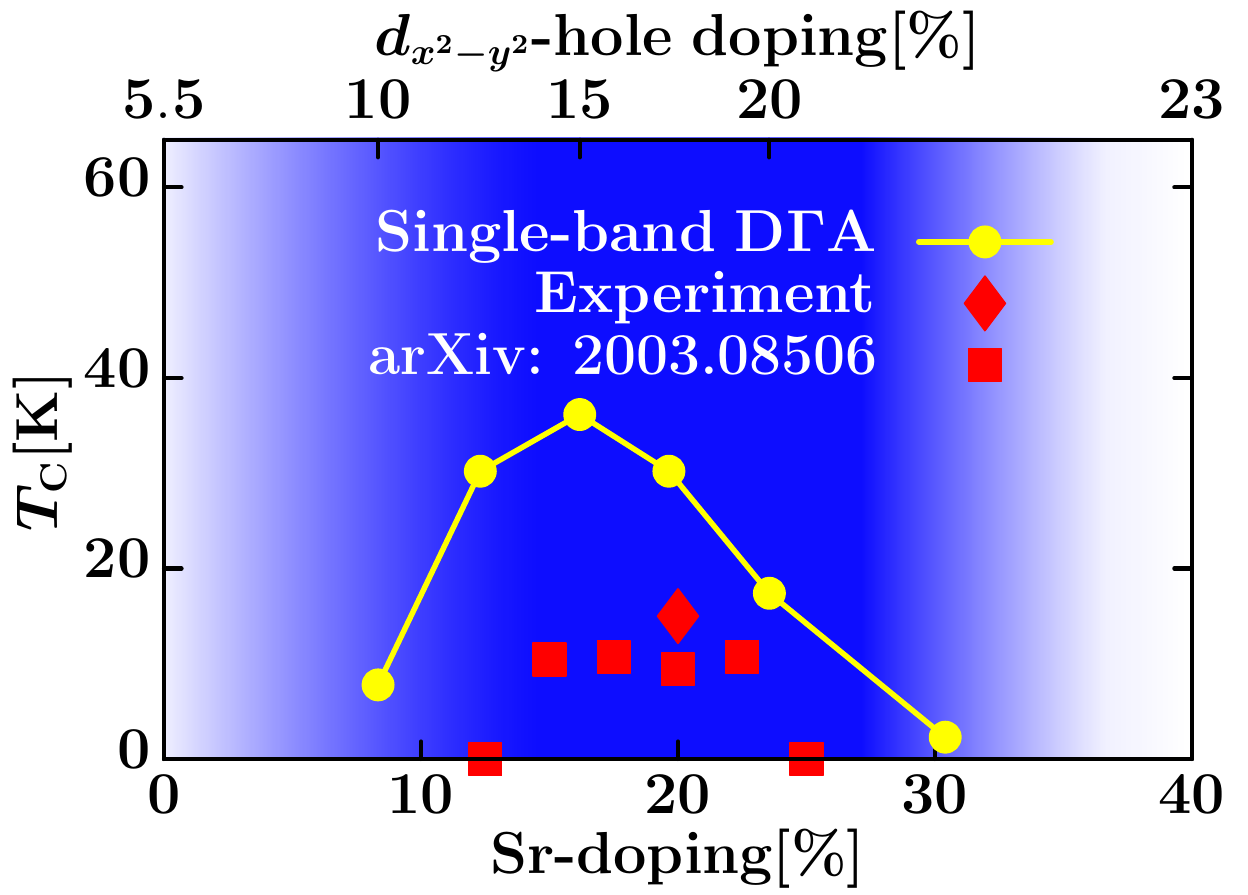}
\par\end{centering}
\caption{Comparison of the theoretical phase diagram at $U=8t$ with the experimental phase diagram of Ref.~\onlinecite{Li2020} (``arXiv: 2003.08506''); ``Experiment'' refers, as before, to Ref.~\onlinecite{li2019superconductivity}. While $U=9t$ in Fig.~\ref{fig:suppl-phase} would be in almost perfect agreement with experiment, we believe the  $T_{\rm C}$ of the present D$\Gamma$A calculation should be slightly larger than the experimental one.}
\label{fig:comp}
\end{figure}

Please consider that our calculation is a prediction of a most difficult quantity to calculate, namely the superconducting $T_{\rm C}$ and its doping dependence, and that we discussed already before that theory should overestimate $T_{\rm C}$,  while the exerimentally observed  $T_{\rm C}$ is likely suppressed by extrinsic contributions such as disorder etc. One may of course most easily compensate this over- and underestimation by changing the $U$ value. Indeed, $U=9t$ of Fig.~\ref{fig:suppl-phase} would be in almost perfect agreement with the experimental phase diagram,  it also was already in excellent with the previously only available data point (Ref.\onlinecite{li2019superconductivity}; ``Experiment'' in Fig.~\ref{fig:comp}).

While $U=9t$ would be essentially on top of the experimental phase
 diagram, we raised some factors that theory overestimates  $T_{\rm C}$  on the
 other hand, and we still think that the true $U$ is $8t$ or in-between $U=8t$ and $U=9t$, and that further improving the calculation and purifying the crystals will eventually converge experimental and theoretical phase diagrams.

We conclude that given this reasonable over- and underestimation we have a very good agreement of the absolute value of  $T_C$ and its doping dependence.

\clearpage

\bibliography{full}
\bibliographystyle{apsrev4-1}